\documentclass[prl,twocolumn,superscriptaddress,showpacs]{revtex4}
\usepackage{amsmath,bpchem}
\usepackage[utf8x]{inputenc}
\usepackage[dvipdfm]{graphicx}

\DeclareMathOperator{\arccosh}{arccosh}

\begin{document}
\title{Anomalous ordering in inhomogeneously strained materials}
\author{Charo I. \surname{Del Genio}}
	\affiliation{University of Houston, Department of Physics, 617 Science \& Research 1, Houston, TX 77204-5005, USA}
	\affiliation{Texas Center for Superconductivity, University of Houston, 202 Houston Science Center, Houston, Texas 77204-5002, USA}
\author{Kevin E. \surname{Bassler}}
	\affiliation{University of Houston, Department of Physics, 617 Science \& Research 1, Houston, TX 77204-5005, USA}
	\affiliation{Texas Center for Superconductivity, University of Houston, 202 Houston Science Center, Houston, Texas 77204-5002, USA}

\date{\today}

\begin{abstract}
We study a continuous quasi two-dimensional order-disorder phase transition
that occurs in a simple model of a material that is inhomogeneously strained
due to the presence of dislocation lines. Performing Monte~Carlo simulations of
different system sizes and using finite size scaling, we measure critical exponents
describing the transition of $\beta=0.18\pm0.02$, 
$\gamma=1.0\pm0.1$, and $\alpha=0.10\pm0.02$.
Comparable exponents have been reported in
a variety of physical systems. These systems undergo a range of different
types of phase transitions, including structural transitions, exciton percolation,
and magnetic ordering. In particular, similar exponents have been
found to describe the development of magnetic order at the onset of the pseudogap
transition in high-temperature superconductors. Their common universal critical
exponents suggest that the essential physics of the transition in all of these
physical systems is the same as in our simple model.
We argue that the nature of the transition in our model is related
to surface transitions although our model has no free surface.
\end{abstract}


\maketitle

\begin{figure}
 \centering
{\includegraphics[width=0.45\textwidth]{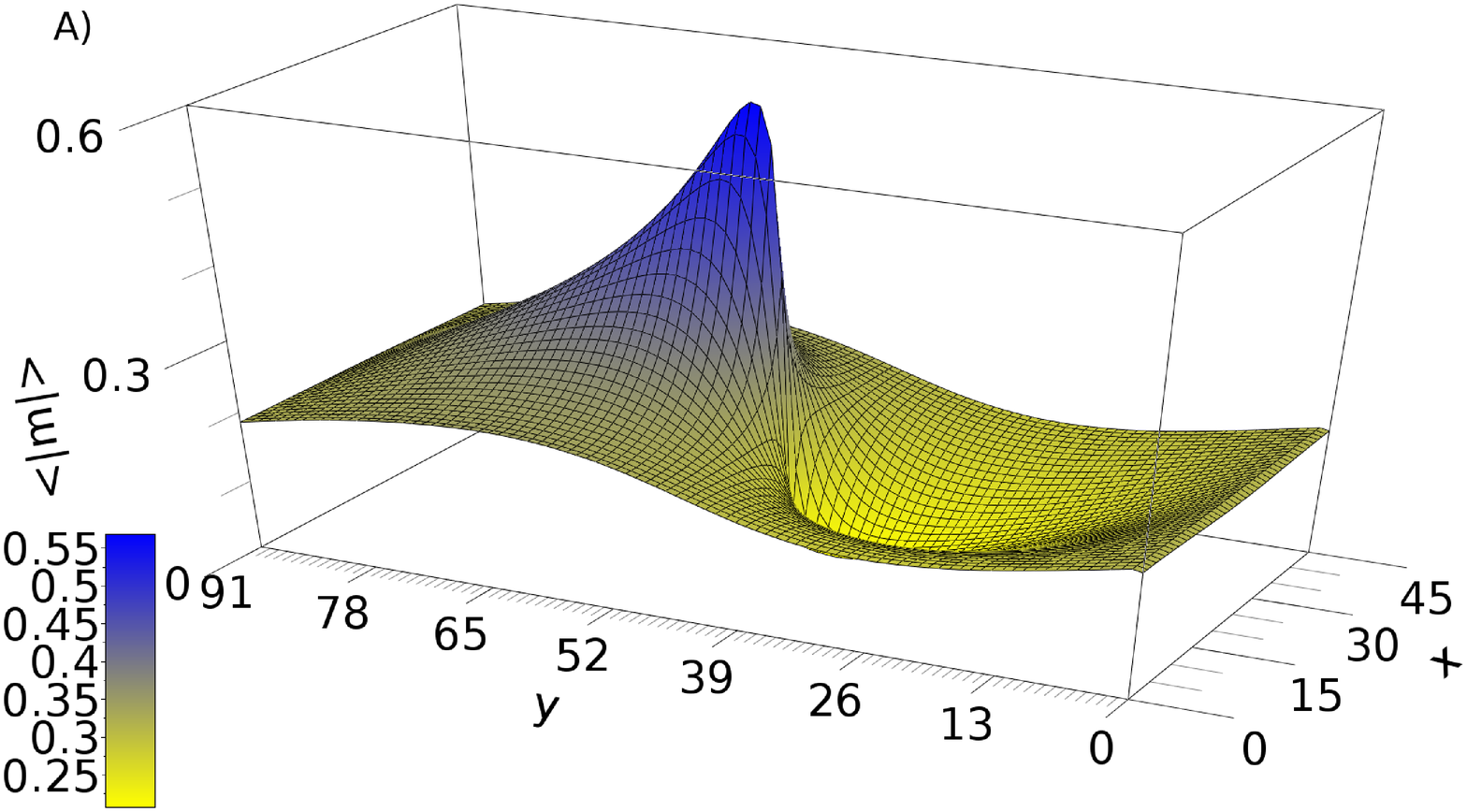}
\vspace{10pt}}
\includegraphics[width=0.45\textwidth]{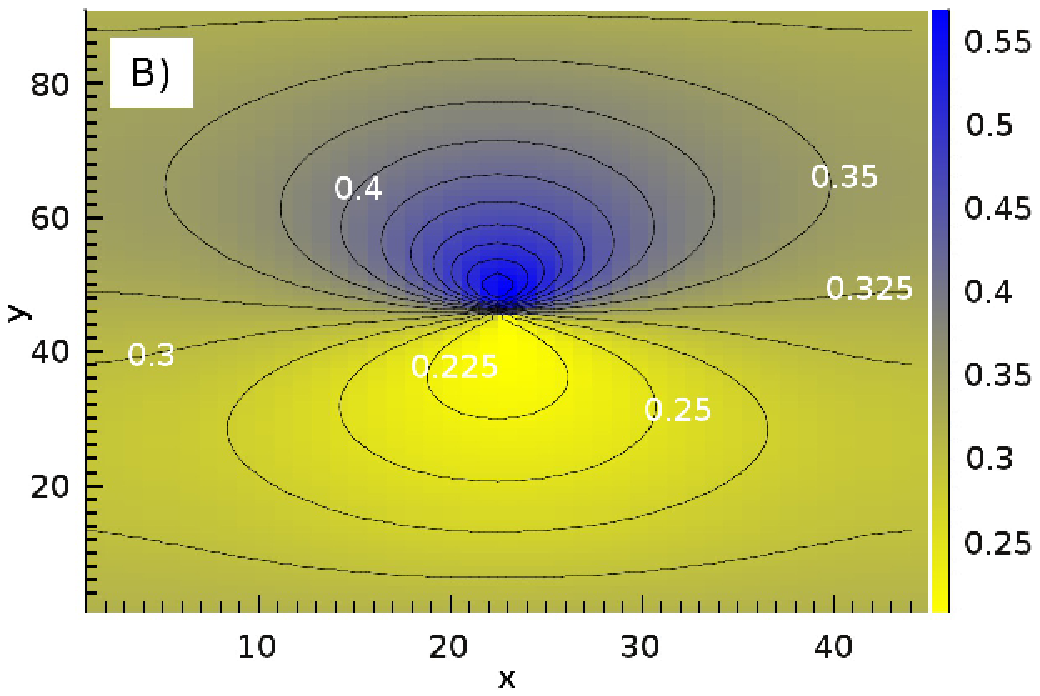}
\caption{\label{Fig1}(Color online) A) Magnetization order
parameter $\langle|m|\rangle$ for the $x$-$y$ cross section, averaged over sites 
in the $z$ direction, of a $45\times91\times40$
system at a temperature of $4.49~k_B$. The order-enhanced and order-suppressed
zones are shown in blue (dark) and yellow (light), respectively. 
B) Contour plot of the same data. The contour lines are in a step
of $0.025$. The innermost line in the blue (dark) order-enhanced region 
corresponds to $0.55$.} 
\end{figure}
\begin{figure}
 \centering
\includegraphics[width=0.43\textwidth]{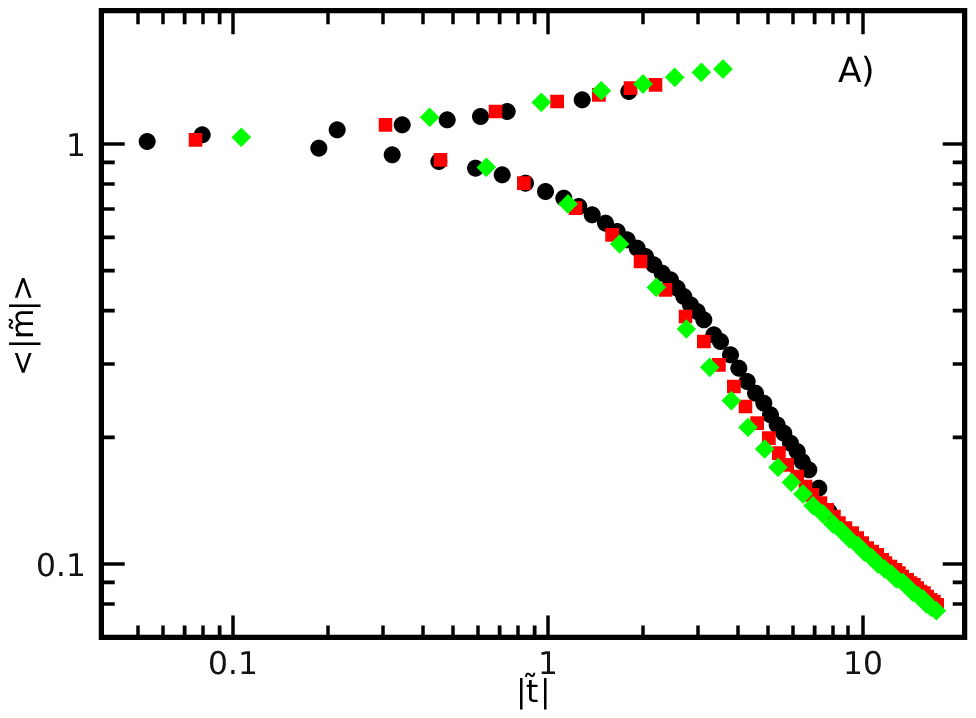}
\includegraphics[width=0.43\textwidth]{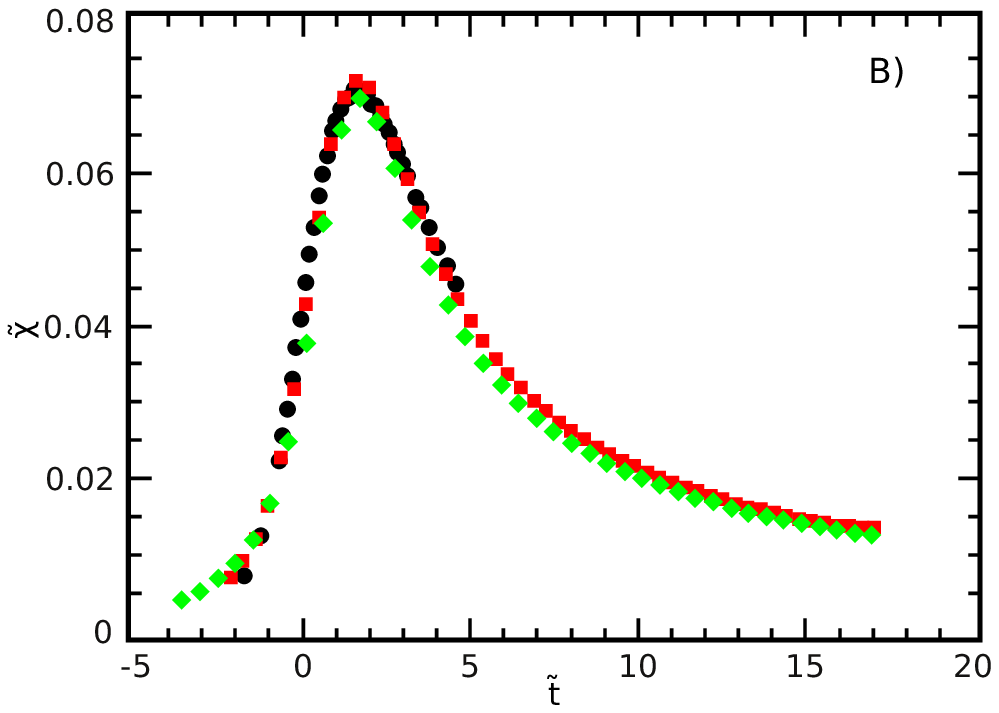}
\includegraphics[width=0.43\textwidth]{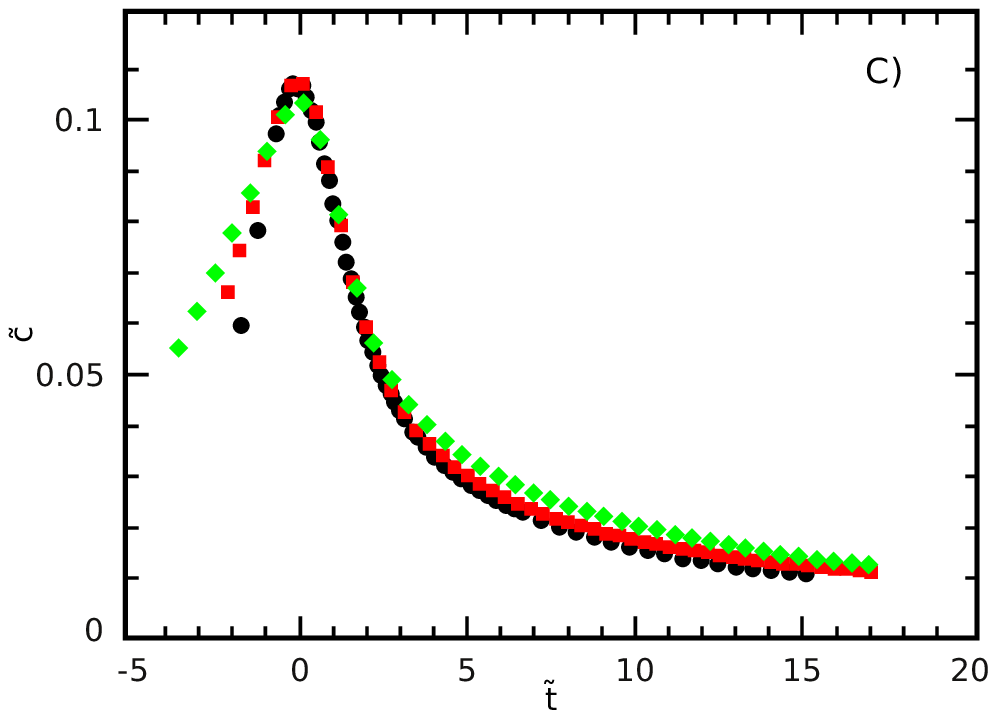}
\caption{\label{Fig2}(Color online) Finite size scaling data
collapses for a quasi two-dimensional layer at the
surface of an ordered cylindrical nucleus.
The black circles correspond to an $x$-$y$ cross section circumference of 50, the red squares to
a circumference of 102 and the green diamonds to a circumference of 200.
The critical exponent $\nu=2.0$ and the critical temperature is $T_c=6.7$.
A) Magnetization scaling function $\langle|\tilde m|\rangle$ vs.~$|\tilde t|$
using the long range order critical exponent $\beta=0.18$.
B) Susceptibility scaling function $\tilde\chi$ vs.~scaled
reduced temperature $\tilde t$ using the critical exponent $\gamma=1.0$.
C) Specific heat scaling function $\tilde c$ vs.~$\tilde t$ using
the critical exponent $\alpha=0.10$.}
\end{figure}
\begin{figure}
 \centering
\includegraphics[width=0.45\textwidth]{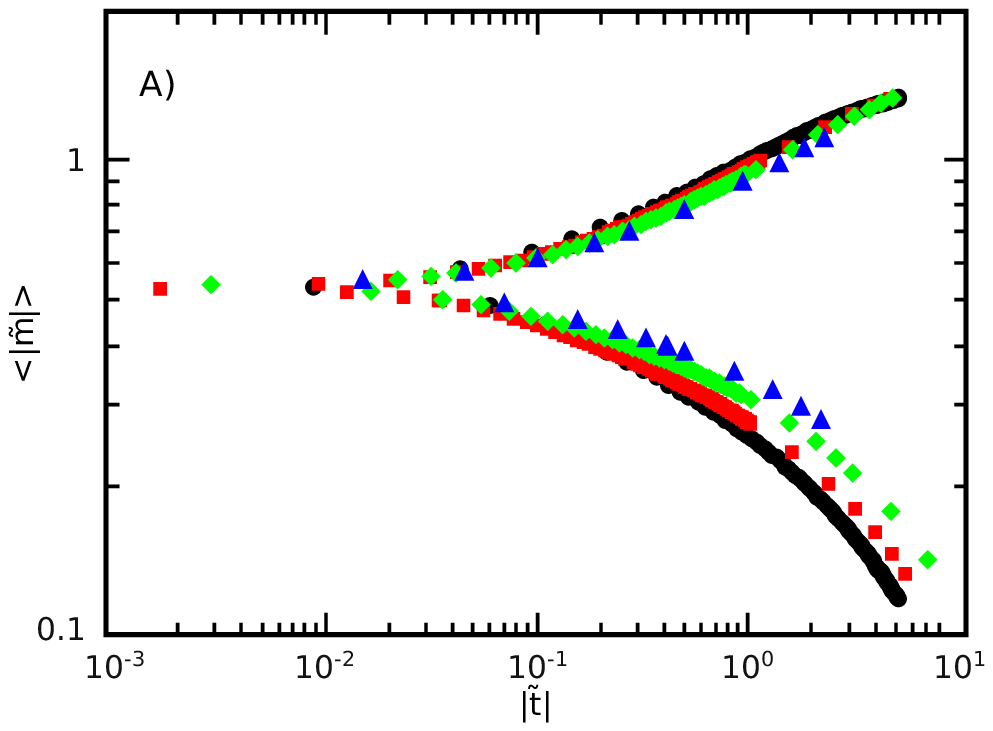}
\includegraphics[width=0.45\textwidth]{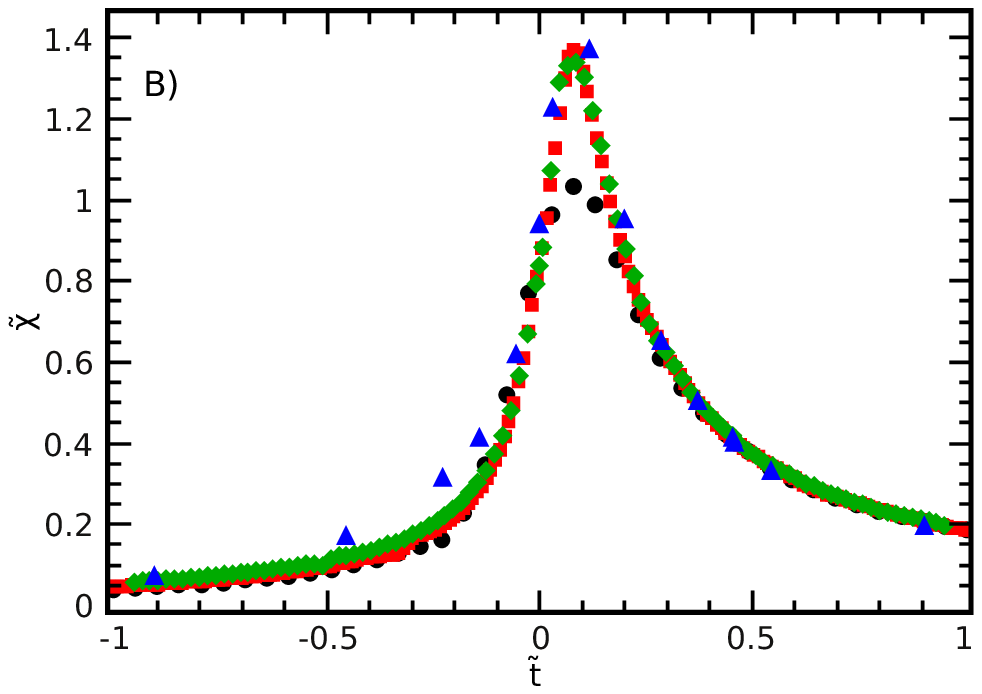}
\caption{\label{Fig3}(Color online) Finite size scaling
data collapses for the whole system.
The black circles correspond to a size of $59\times23\times13$, the red
squares to a size of $109\times43\times25$, the green diamonds to a size
of $205\times83\times50$ and the blue triangles to a size of $417\times167\times101$.
The critical exponent $\nu=2.0$ and
the critical temperature is $T_c=4.50$.
A) Magnetization scaling function $\langle|\tilde m|\rangle$ vs.~$|\tilde t|$
using
the long range order critical exponent $\beta=0.18$.
B) Susceptibility scaling function $\tilde\chi$ vs.~scaled
reduced temperature $\tilde t$
using
the critical exponent $\gamma=1.1$.}
\end{figure}

Real solids are commonly in a strained state. This can be due to a variety
of reasons, ranging from forces applied upon them to the presence of structural
defects, to ongoing phase transformations. 
Such strains affect the ordering processes of the materials~\cite{Del10,Del09,Li10,Cah06,Li05,Dub79}.
Therefore, understanding the extent of these effects is important. In this
Communication, we study the continuous order-disorder phase transition in a model
of a strained material. The strain field we consider results from a ``wall
of dislocations'', that is, a linear array of parallel edge dislocation lines.
This particular arrangement of defects is relatively common
in  crystals, as it often occurs because of surface treatments. The resulting strain
is inhomogeneous, and order develops inhomogeneously in the material,
with ordered regions growing in quasi two-dimensional layers around a
central cylindrical rod-shaped nucleus~\cite{Del10}. Each layer orders at a
different critical temperature. In order to study the critical behavior of
this process, we consider a mesoscopic spin model in which the coupling between
spins reflects the strain field induced by the dislocation walls. Performing
several simulations of systems with different sizes and using finite size scaling,
we are able to measure the critical exponents characterizing the transition.
The critical exponents found 
are comparable with exponents that have been measured experimentally
in a variety of materials, and for different types of
transitions~\cite{Del10,Del09,Tre98,Sch87,Tak09,Lom96,Bag93,Gau88,Bon77,Yel74}.
Notably, similar critical exponents have recently been measured for the magnetic
ordering transition that accompanies the onset of the pseudogap state in high
$T_c$ superconductors~\cite{Son09,Son08,Moo08,Li08}. These exponents
are also compatible with those found in 
multicritical surface transitions~\cite{Bin84},
although in our case the exponents describe bulk measurements.

Assuming that atoms interact
more strongly where they are pushed closer together and more weakly where they
are pulled apart, a phenomological model that captures the effect of strain 
on ordering due to a dislocation line can
be constructed~\cite{Del10}. Assuming the defects are arranged in walls extending in the
$y$ direction with the lines parallel to $z$, it is found that the local relative
critical temperature change $\tau_c(\vec{r})$ is
\begin{equation}\label{LCTC}
 \tau_c\left(\vec r\right)\equiv\frac{T'_c\left(\vec r\right)-T_c}{T_c}\propto\frac{b}{2l}\frac{1-2\nu}{1-\nu}\frac{\sin\left( \frac{2\pi y}{h}\right) }{\cosh\left( \frac{2\pi x}{h}\right) -\cos\left( \frac{2\pi y}{h}\right) }\:,
\end{equation}
where $\vec r$ is the normal vector pointing from the closest dislocation
line, $b$ is the magnitude of the Burgers vector, $l$ is the unit of length
used, $\nu$ is Poisson's ratio, $h$ is the local average distance between
defects, $T'_c\left(\vec r\right)$ is the local transition temperature and 
$T_c$ is the transition temperature for a defect-free crystal. 
This results in inhomogeneous ordering in which
ordered regions nucleate and grow in the vicinity
of the dislocation lines via the addition of quasi~2-D layers around nuclei
with the shape of narrow cylindrical rods~\cite{Del10}.
Here we study the universal critical scaling properties of this ordering
process.


To identify the essential physics that controls the
scaling properties of this ordering behavior,
we studied a zero-field 3D~Ising model on a simple
cubic lattice with periodic boundary conditions and nonconstant coupling $J_{ij}$ between
nearest neighbor spins $i$ and $j$. The Hamiltonian is
\begin{equation}\label{Ham}
 \mathcal H=-\sum_{\left\langle ij\right\rangle}J_{ij}s_is_j\:,
\end{equation}
where $s_i=\pm 1$ is the value of the $i^\mathrm{th}$ spin
and $\left\langle ij\right\rangle$ indicates sum over the
nearest neighbor spins on the lattice.
The spins simply represent the
state of local order. The value of the coupling
$J_{ij}$ is chosen in order to reflect the strain field
giving rise to Eq.~\ref{LCTC} in the following way.
First note that in a ``regular'' Ising model, with constant
coupling $J_0$, the critical temperature is proportional to
the coupling constant:
\begin{equation}\label{TcOrig}
 J_0=\frac{T_c}{a}\:,
\end{equation}
where $a$ is some proportionality constant. Also, from Eq.~\ref{LCTC}
it follows that, given a $\tau_c\left(\vec r\right)$,
\begin{equation}\label{TcPrime}
 T_c'\left(\vec r\right)=T_c\left(1+\tau_c\left(\vec r\right)\right)\:.
\end{equation}
Therefore, from Eqs.~\ref{TcOrig} and~\ref{TcPrime},
the parts of the system that become critical at a given
temperature $T_c'$ are those that have a coupling
\begin{equation*}
J(\vec{r})
 =\frac{T_c'\left(\vec r\right)}{a}=\frac{T_c}{a}\left(1+\tau_c\left(\vec r\right)\right)=J_0\left(1+\tau_c\left(\vec r\right)\right)\:.
\end{equation*}
Thus, given the arbitrarity of $J_0$ and of the other proportionality constants,
we set
\begin{equation*}
 J\left(\vec r\right)=1+\frac{\sin\left( \frac{2\pi y}{h}\right) }{\cosh\left( \frac{2\pi x}{h}\right) -\cos\left( \frac{2\pi y}{h}\right) }\:,
\end{equation*}
where we take $h$ to be the size of the system in the $y$ direction.
To reproduce the effects of the strain of a wall, we use the
above expression only for the coupling between spins in the $x$ and $y$
directions, while we set the coupling of the spins along $z$ at 1.
The simulated systems contained a single dislocation line in the center.
The replicas due to the periodic boundary conditions used effectively
turned it into a wall of lines. Notice that while the strain field
due to a single dislocation line is long-range, the one due
to a wall is short-ranged~\cite{Del10}. However, the field of a wall maintains the
dipole-like nature of the field of a single line, with the effect of
promoting the order on one side of the system, while suppressing it on the
other. The order parameter in our simulations was given by the ensemble averaged
absolute
value of the magnetization per spin:
\begin{equation*}
\left\langle \left|m\right| \right\rangle = \frac{1}{N}\left|\sum_is_i\right|\:,
\end{equation*}
where $N$ is the total number of spins.

Using the Wolff algorithm~\cite{Wol89}, which is a cluster flipping
algorithm~\cite{Swe87}, we performed extensive Monte~Carlo simulations
of this model. 
The cylindrical ordered regions grow with
decreasing temperature as the surfaces of the cylinders order in a
fashion consistent with earlier predictions~\cite{Del10}. 
Figure~\ref{Fig1}
shows the order parameter in an $x$-$y$ cross section of a $45\times91\times40$
system, averaged over $z$, at a temperature of $4.49$ in units of the
Boltzmann constant $k_B$. As anticipated, order is increasingly enhanced
with proximity to the dislocation line on one half of the system. On
the other half, instead, order is increasingly suppressed. Also notice
that the contour lines closely follow the predicted shape, shown in
Fig.~6 of Ref.~\onlinecite{Del10} and computed by numerically solving
the following parametric equations for a particular value of $\tau_c$:
\begin{gather*}
y\left(x\right)=\frac{h}{\pi}\arctan\left\lbrace\frac{\pi\pm\sqrt{\pi^2+\tau_c^2\left[1-\cosh^2\left(\frac{2\pi x}{h}\right)\right]}}{\tau_c\left[\cosh\left(\frac{2\pi x}{h}\right)+1\right]}\right\rbrace\:;\\
x\left(y\right)=\pm\frac{h}{2\pi}\arccosh\left\lbrace\frac{2\pi\tan\left(\frac{\pi y}{h}\right)+\tau_c\left[1-\tan^2\left(\frac{\pi y}{h}\right)\right]}{\tau_c\left[1+\tan^2\left(\frac{\pi y}{h}\right)\right]}\right\rbrace\:.
\end{gather*}

We find that the ordering occurs via a continuous transition.
To measure the critical exponents,
we simulated systems of different sizes and estimated their
values using finite size scaling~\cite{New99}. The observables
measured were the magnetization order parameter and the ensemble averaged total
energy, given by Eq.~\ref{Ham}. From the fluctuations of magnetization
and energy we also calculated the magnetic susceptibility $\chi$
and the specific heat $c$. The measurements were taken at the
same time over the entire system and over an arbitrarily chosen
quasi two-dimensional layer, corresponding to a fixed, chosen value
of $\tau_c$. For each value of the temperature
we took ensemble averages over a number of system updates between
$10^6$ and $10^8$. The whole system sizes were $59\times23\times13$,
$109\times43\times25$, and $205\times83\times50$, while the circumferences
of the $x$-$y$ cross sections of the quasi two-dimensional layers measured
were 50, 102 and 200, corresponding to $\tau_c=0.9$. The sizes of the
systems in the $x$ direction were chosen so that the coupling between spins
was within $10^{-6}$ of unity at the boundaries.

To perform data collapses using finite size scaling, we define the scaled
reduced temperature $\tilde t$ as
\begin{equation*}
 \tilde t\left(t\right)=L^{1/\nu}t\:,
\end{equation*}
where $L$ is the length of the largest dimension
of the system considered, which in our
case corresponds to the length in the $x$ direction, $\nu$ is the correlation
length critical exponent and $t\equiv\frac{T-T_c}{T_c}$ is the reduced
temperature. With this definition of $\tilde t$, the scaling functions for
the order parameter, the magnetic susceptibility and the specific heat
are, respectively,
\begin{gather*}
 \langle|\tilde m|\rangle\left(\tilde t\right)=L^{\beta/\nu}\langle|m|\rangle\left(\tilde t\right)\:,\\
 \tilde\chi\left(\tilde t\right)=L^{-\gamma/\nu}\chi\left(\tilde t\right)\:,\\
 \tilde c\left(\tilde t\right)=L^{-\alpha/\nu}c\left(\tilde t\right)\:,
\end{gather*}
where $\beta$, $\gamma$ and $\alpha$ are the corresponding critical
exponents. The data collapses for the quasi two-dimensional layer,
shown in Fig.~\ref{Fig2},
allow estimates of the critical indices of $\beta=0.18\pm0.02$,
$\gamma=1.0\pm0.1$, $\alpha=0.10\pm0.02$ and $\nu=2.0\pm0.1$,
with a critical temperature
of $6.7\pm0.2$. The errors were conservatively estimated as the range
over which a reasonable scaling collapse was achieved.

Similarly, the measurements of the whole systems, whose data collapses are shown
in Fig.~\ref{Fig3}, allow the values of the critical exponents to be estimated
as $\beta=0.18\pm0.02$, $\gamma=1.1\pm0.2$ and $\nu=2.0\pm0.25$, with
a critical temperature of $4.50\pm0.05$. We could not
produce a good scaling collapse for the specific heat.
Note that we get essentially the same exponents for the whole
system that we do for the quasi two-dimensional layer. This reveals
that the nature of the transition of the whole system is essentially the same as
that of a quasi two-dimensional layer.
At any given temperature, there is a part of the system that is critical.
The biggest of these parts corresponds to the measured critical temperature
for the whole system. Also notice that the susceptibility for the smallest
system does not scale well near the peak, presumably due to finite size effects.
The size of the error bars on the data shown in Figs.~\ref{Fig2} and~\ref{Fig3}
is substantially smaller than the size of the symbols.

The exponents characterizing the transition are compatible with those
corresponding to the, so-called, ``special'' multicritical point in surface critical
phenomena. In particular, the value $\beta=0.18$ was reported in Refs.~\cite{Bin84}
and~\cite{Lan90} and is consistent with prior theoretical calculations based on scaling~\cite{Ree81,Bin84}.
Also, the measured mean-field value of the exponent $\gamma=1$ is expected at the
multicritical point~\cite{Bin74}. 
Furthermore, 
using the ``bulk'' 3d-Ising value for the
critical exponent $\nu=0.632$ in the scaling
laws, as in Ref.~\cite{Bin84},
the hyperscaling relation 
predicts $\alpha = 0.104$, which is compatible with the
one we measured.
Note, however, that while these previous studies
considered systems with an actual surface, our model does not have free layers.
In fact, the quasi two-dimensional layers whose ordering we studied are in the
midst of of the system. Nevertheless, the ordering in our system does occur in
a quasi two-dimensional layer at the surface of the already ordered region.

Similar exponents have also been measured for a number of different types
of transitions in a variety of physical systems, ranging from structural transitions,
to the percolation of excitons in polymeric matrices, to magnetic order in
frustrated materials~\cite{Del10,Del09,Tre98,Sch87,Tak09,Lom96,Bag93,Gau88,Bon77,Yel74}.
In particular, as mentioned earlier, there have been recent observations
of magnetic ordering at the onset of the pseudogap transition in high-$T_c$
superconductors in which similar critical exponents have been 
measured~\cite{Son09,Son08,Moo08,Li08}.
Given the scale invariant nature of critical phenomena, the
fact that the phase transition in our model apparently has the same set of critical
exponents suggests that the essential physics is the same in both systems,
and that our results may be relevant to the open question of the nature of the
pseudogap state itself. Intriguingly, recent experiments have shown that the onset
of the pseudogap state is accompanied by local modulations of atomic displacement
that generate significant inhomogeneous 
strains~\cite{Isl02,Isl04}.
This suggests that, like the quasi two-dimensional ordering process
we have considered, the pseudogap transition occurs because of inhomogeneous strain.

Assuming this is true 
and noting that
the pseudogap transition precedes the onset of 
high-$T_c$ superconductivity~\cite{Moo08},
it appears that some strain is required for the development of
high-$T_c$ superconductivity.
However,
strain is also known to adversely affect
superconductivity~\cite{Cal04,Bus04,Zhu03,Jan03}
and too much strain supresses it altogether~\cite{Li08}. 
The optimal doping
concentration of the high-$T_c$ superconductor \BPChem{YBa\_2Cu\_3O\_{7-$\delta$}}
occurs at only $\delta\approx0.08$. Such a small deviation from an exact
stoichiometry presumably introduces 
enough strain to cause a pseudogap transition
while causing only minor adverse effects.
This supports the idea that the pseudogap state is a physically direct precursor to
superconductivity, even though its cause competes with it, 
consistent with some of the original ideas concerning the mechanism of
high-temperature superconductivity~\cite{Eme87,Eme95}.

\begin{acknowledgments}
The authors are grateful to Simon~C. Moss for many helpful discussions.
This work was supported by the NSF through grant No.~DMR-0908286 and by
the Texas Center for Superconductivity at the University of Houston (T${}_\mathrm{c}$SUH).
\end{acknowledgments}

\end{document}